\documentclass[namedreferences]{solarphysics}
\usepackage{amsmath}
\usepackage[hyperref,optionalrh,solaromanenum]{spr-sola-addons} 
\usepackage{graphicx}                    
\usepackage{color}                       
\usepackage{epstopdf}
\AppendGraphicsExtensions{.tif}

\usepackage{solphys_macros}

\usepackage{xcolor}
\usepackage{soul}

\newcommand{\Fexix}{\mbox{Fe\,\textsc{xix}}}
\newcommand{\Fexxiii}{\mbox{Fe\,\textsc{xxiii}}}
\newcommand{\Fexxiv}{\mbox{Fe\,\textsc{xxiv}}}

\begin{document}

\begin{article}

\begin{opening}

\title{Chromospheric evaporation by particle beams in multi-dimensional flare models}

%
\author[addressref={1},corref,email={malcolm.druett@kuleuven.be}]{\inits{}\fnm{Malcolm Keith }\lnm{Druett}\orcid{0000-0001-9845-266X}}
\author[addressref={1},corref,email={}]{\inits{}\fnm{Wenzhi }\lnm{Ruan}\orcid{}}
\author[addressref={1},corref,email={}]{\inits{}\fnm{Rony }\lnm{Keppens}\orcid{0000-0003-3544-2733}}

%

\address[id={1}]{Centre for mathematical Plasma Astrophysics, Department of Mathematics, KU Leuven, Celestijnenlaan 200B, B-3001 Leuven, Belgium}

\begin{abstract}
   Evaporation of chromospheric plasma by particle beams has been a standard feature of models of solar flares for many decades, supported both by observations of strong hard X-ray bremsstrahlung signals, and detailed 1D hydrodynamic radiative transfer models with near-relativistic electron beams included. However in multi-dimensional models, evaporation, if included, has only been driven by heat conduction and by the impact and reflection of fast plasma outflows on the lower atmosphere.
   Here we present the first multi-dimensional flare simulation featuring evaporation driven by energetic electrons.
   We use a recent magnetohydrodynamic model that includes beam physics, but decrease the initial anomalous resistivity to create a gentler precursor phase, and improve on the dynamic resistivity treatment that determines where beams are injected.
   Beam-driven evaporation is achieved. The relevant factors are thermal conduction and electron beams, with the beam electrons more than doubling the kinetic energy flux, and adding 50\% to the upward mass from the chromosphere.
   These findings finally pave the way for integrating detailed 1D flare modelling within a self-consistent 2D and 3D context. The beam fluxes from these self-consistent models can be used to directly compare multi-dimensional results with those from the externally injected beam fluxes of 1D models, as well as understand further evaporation-driven phenomena relating to beams of particles.
\end{abstract}

%

\end{opening}

%
\section{Introduction} \label{sec:intro} 

\subsection{Observations of chromospheric evaporation} \label{sec:obs} 

Chromospheric material is typically considered to be at temperatures between $6,000$~K and $12,000$~K with plasma number densities in the range $10^{9}$ (generally in the upper regions) to $10^{13}$ cm$^{-3}$ near the base \citep{1993FontenlaFALC, 2019MolnarChromoTemp}. The chromosphere is a complicated layer with 3-dimensional (3D) morphology that is situated between the photosphere and the transition region. In flares, chromospheric evaporation is due to dramatic heating and upflow of this chromospheric material into the corona. It is considered to be a primary mechanism for filling flare loops with hot, relatively dense material that is then responsible for the strong extreme-ultraviolet (EUV) emissions detected.

Such evaporations have been reported above flare ribbons since EUV and X-ray observations of flares became available \citep{1999Czaykowska, 2004Veronig, 2006Milligan, 2009MilliganEvap, 2011Aschwanden, 2017Druett}. The locations of the footpoints of these evaporations are commonly associated with strong HXR Bremsstrahlung \citep{2006Milligan, 2009MilliganEvap, 2017Druett} despite the low resolution imaging of HXR instruments. This emission is associated with the breaking energy of high energy electron beams, decelerating due to collisions with thermal plasma. 
\citet{1968Neupert} describe how the time derivative of the soft x-ray (SXR) signal during the increasing phase of flares is well correlated with the hard x-ray (HXR) sources. This ``Neupert effect" is also explained by the HXR signal being associated with the chromospheric footpoints of flare loops which undergo evaporation flows and thereby provide a source of thermal SXR in the looptops.
Chromospheric evaporation remains a key topic of investigation, as demonstrated by recent publications \citep{2022Kerr, 2023Kerr, polito_ribbons_2023}.

\subsection{Evaporation in 1D models} \label{sec:1d} 

1-dimensional (1D) solar flare models have included the effects of beam electrons heating a ``thick target" (i.e. losing all their kinetic energy) in the lower atmosphere since the 1970s \citep{1972Syrovatskii, 1978Emslie, 1981Somov, 1983Duijveman, 1985Fisher, 1987Canfield}.
This approach has advanced our understanding of flare phenomena that has continued through to modern models. A popular 1D flare model family is RADYN \citep{2005Allred, 2015Allred, 2017SimoesFlareContinuum, polito_ribbons_2023, 2023CarlssonFCHROMA} which, in their latest form, include beam electron return currents \citep{1995Zharkova} and distributions determined by solving the Fokker-Planck equation \citep{2020Allred}. Other detailed 1D hydrodynamic radiative transfer codes have also contributed strongly to the field, for example FLARIX \citep{1995Heinzel, 2019FLARIX}, and HYDRO2GEN \citep{1993Zharkova, 2017Druett, 2018Druett, 2019Druett}. 

These detailed 1D models have provided great insight with regard to interpreting spectral signatures of white light flares \citep{2018Druett, 2017SimoesFlareContinuum}, continuum emissions \citep{2017SimoesFlareContinuum, 2019Druett, 2023McLaughlin}, chromospheric condensations \citep[cool, downward moving plasma at the footpoints of flare loops, ][]{2017Druett, 2017Kowalski, 2020Graham, 2022Kowalski} and evaporations \citep{2017Druett, 2019Druett, 2022Kerr, 2023Kerr, polito_ribbons_2023}. Their findings have also been applied to gain insight for stellar flare processes \citep{2017KowalskiStellar, 2018Kowalski, 2019Kowalski, 2023Kowalski}.

Thermal flares which use thermal conduction as the principal mode of evaporation \citep{1980Nagai, 1982Somov}, and other energy transport mechanisms such as Alfv\'en waves have been suggested \citep{2008Fletcher}, but more intense study to decide on their relative importance is still needed.

\subsection{Evaporation in 2D and 3D models} \label{sec:multid} 

Recent papers reported full 3D MHD flare simulations. The MURaM Code \citep{2005muram}, has been used to study the build-up and release of magnetic energy in flares  \citep{2019Cheung_flare, 2023Rempel_flare} and a recent chromospheric extension to the code could be a powerful tool for future flare simulations \citep{2022Przybylski}. None of the published MURaM flare models incorporate electron beam physics, and the radiative MHD treatments lead to realistically looking, but somewhat excessively hot coronal evolutions, where self-consistent convectively-driven flux emergence and reshuffling gives rise to highly complex atmospheric evolutions.

\cite{2014GuoSADs}, and subsequently \cite{2022Shen3DFlare} performed 3D flare simulations of the impulsive phase in a standard flare model setup, again with focus on energy release in the corona which replicated and interpreted the Rayleigh-Taylor-like supra-arcade downflows (SADs) observed above flare looptops. 
Previously SADs had been used to infer reconnection rates, by using their speeds as velocities of reconnection jet outflows, however this work demonstrated that these features should be interpreted as flows between the termination shock of the reconnection jet and the hot flare looptops. Associated papers by \cite{2022KongXRay, 2022KongAcceleration} have also modelled electron transport in detail within a multi-dimensional setting to simultaneously explain HXR sources in the looptops and the footpoints of flares. Most recently, \cite{2023Ruan3D} pointed out that in 3D standard flare model settings, the Kelvin-Helmholtz (KH) instability develops in flare loop top regions before any Rayleigh-Taylor effects set in. This KH process causes turbulence and its Alfv\'enic propagation down to the footpoint regions. Their pure 3D MHD approach produced results that correspond well with observed non-thermal velocity evolutions in observations, but did not address evaporation physics.

\cite{2018Bakke, 2020Frogner, 2023Frogner} implemented fast electron particle beams in the Bifrost code \citep{2011Gudiksen} and were able to study small scale reconnection events filled within a full 3D volume. Actual evaporation dynamics is yet to be demonstrated, as the published works either did not include a chromospheric response, or focus on the enormous technical challenges associated with incorporating beams along dynamical fieldlines.

\subsection{Presenting the first chromospheric evaporation by beam electrons in a multi-dimensional model} \label{sec:ourmodel} 

Despite the many discoveries of multi-dimensional flare models, none has reproduced chromospheric evaporation via beams of particles, as is ubiquitous in 1D simulations. In this paper we present the first model that is capable of this. Our models are self consistent and therefore do not specify the electron fluxes \textit{a priori} in individual beams, which could be useful for direct comparison with results from 1D modelling. Detailed 1D flare models are embedded within our simulation domain and can be used in the future to compare and test predictions from different 1D models within a multi-dimensional context, although our multi-dimensional models do not yet include detailed non-local thermodynamic equilibrium radiative transfer, which 1D modelling shows is important for the energy balance of the chromosphere, and thus chromospheric evaporation \citep{2005Allred, 2015Allred, 2018Druett, 2019FLARIX}. We can also alter parameters such as the resistivity and background magnetic field strength in order to calibrate our resultant beams against those from 1D models. One could also remove the self-consistency of the model for testing specific beam parameters within a multi-dimensional context.

The reproducable, multi-dimensional standard flare setup models presented here are implemented in the Message Passing Interface Adaptive Mesh Refinement Versatile Advection Code \citep[MPI-AMRVAC,][]{2012KeppensAMRVAC, 2014PorthAMRVACSolar, 2018XiaAMRVACSolar,2023KeppensAMRVAC}. In atmospheric structure and evolution they are descendants of the simulations by \cite{2001Yokoyama}, whose 2D models consist of a weak bipolar field region (``weak" in the context of reported strengths of photospheric magnetic field in flares) with a patch of anomalous resistivity placed in a current sheet in the corona. This resistivity triggers reconnection and converts free magnetic energy to other forms such as Ohmic heating via a non-conservative term in the induction equation \citep[see][]{2023DruettAMRVAC}. Reconnection jets flow outwards from this coronal resistivity patch in the upward and downward direction typically taking a lobster-claw form \citep{2011_Zenitani_reconnection_outflows}, and sometimes including plasmoids due to the tearing instability. When this outflow impacts the lower atmosphere there is strong reflection causing a hot explosive chromospheric evaporation as well as a cooler downward chromospheric condensation due to the impact. Subsequent to this impact, evaporation flows are sustained by thermal conduction. The resultant evaporations were studied in detail by \cite{2015Takasao}, and their effects on the flare looptops and coronal emission has been presented also in \cite{2018RuanKHI, 2019RuanEUVXRay, 2020RuanFlare, 2021RuanRain} including, recently, in 3D with synthetic observables presented \citep{2023Ruan3D}.

\begin{figure}
    \centering
    \includegraphics[width=\textwidth]{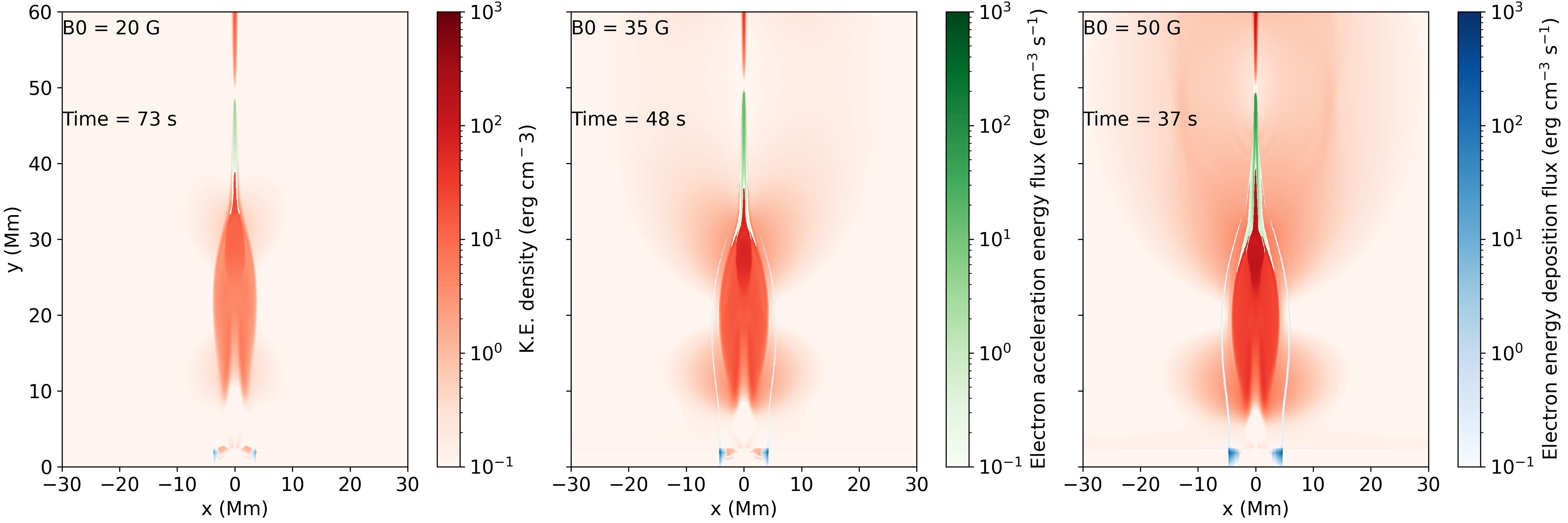}
    \caption{Kinetic energy signatures due to energetic electron beams at the loop footpoints of flare simulations from \cite{2023DruettAMRVAC}. Kinetic energy densities are shown in the background in a red colour scale. The electron acceleration sites and energy densities are identified via overplotted green colours, and the electron beam deposition sites near the footpoints and their associated energy densities are shown using blues, with tails to white at the lower end of the energy scale highlighting lower deposition along the active fieldlines. These quantities are shown just before the reconnection outflow jet (see the red, lobster claw-shape structure in the corona) impacts the dense lower atmosphere. Simulations with background magnetic field strengths of $B_0 = 20, 35$, and $50$~G are shown in the panels from left to right. The evaporation signatures can be seen via the red kinetic energy density at the bases of the models. The signatures neatly cover the region swept over by the beams of energetic electrons. They are subsequently swamped by the explosive evaporation occurring due to reflection of the impact of the reconnection outflow jets on the lower atmosphere. Note that the experiments with different background field strengths evolve at different rates, but the electrons in these simulations were all ``switched on" at $t=31.2$~s, which explains why the beam electrons for the experiments with lower $B_0$ sweep outward from positions initially closer to the polarity inversion line at $x=0$.}
    \label{fig:KEsignatures}
\end{figure}

Another key development reported in \cite{2020RuanFlare} was the implementation of 1D energetic particle beams within the 2.5D models. These used the Ohmic heating term from previous models as an energy reservoir for direct current acceleration of electrons when the electron drift velocity exceeds a threshold value. Our accompanying paper \citep{2023DruettAMRVAC} presents a parameter study of these self-consistent beam-MHD models. In the pioneering study from \cite{2020RuanFlare}, there was insufficient energy carried in the beams to cause beam-driven evaporation, although our parameter survey shows that in certain cases there were signs of beams influencing the kinetic energy densities near the footpoints of the flares before the reconnection outflow jet impacted the lower atmosphere. This is seen clearly in Figure \ref{fig:KEsignatures}, where we show for three different field strengths the kinetic energy distribution right before the impact, and the panels give detailed views on the current beam location (white lines), and both beam acceleration (green) and beam energy deposition (blue) sites.
The investigation presented in this work demonstrates the first chromospheric evaporation by beam electrons in multi-dimensional flare modelling.

\section{Methods} \label{sec:method} 

In this work we focus on the challenge of realistically replicating chromospheric evaporation by electron beams, within a self consistent multi-dimensional model. 
One could also follow a non-self-consistent approach to solar flares by introducing additional energy to account for a reservoir that is stored in, and released from, a complex braided magnetic flux rope running orthogonal to the plane of the 2.5D. We will focus on the self-consistent approach.

Therefore, we seek implementations that will realistically and efficiently channel a greater proportion of the energy from the reconnection process into accelerated particles. This approach leaves several paths open, all of which relate to the modelling of resistivity, $\eta$. In \cite{2020RuanFlare} the anomalous resistivity is modelled in two phases. Firstly there is an initial phase $t<t_\eta$ where

\begin{equation}
    \eta (x,y,t<t_\eta) =
        \begin{cases}
            \eta_0 \left[ 2 \left( \frac{r}{r_\eta} \right)^3 - 3 \left( \frac{r}{r_\eta} \right)^2 + 1 \right] & r \leq r_\eta \\
            0 & r > r_\eta
        \end{cases}
        .
        \label{eq:eta_1}
\end{equation}
This defines a circular patch of resistivity with radius $r=r_\eta$, which is monotonically decreasing in value from $\eta=\eta_0$ to $\eta=0$ outside this disk. We centre this patch at a height of 50~Mm and above the polarity inversion line at $x=0$~Mm, with $r_\eta=2.4$~Mm 

Then from $t=t_\eta$ onward, the anomalous resistivity only occurs in regions where the drift velocity of the electrons, $v_d$, exceeds the threshold velocity for acceleration $v_c$. In the rest of this investigation we keep the threshold fixed as defined in \citet{2020RuanFlare}, i.e. $v_c=1000 u_v$, with $u_v = 128 $km~s$^{-1}$ which is the unit of velocity used for non-dimensionalisation of the experiment. 
\begin{equation}
    \eta (x,y,t \geq t_\eta) =
        \begin{cases}
            0 & v_d > v_c \\
            \mathrm{min} \left\{ \alpha \left( \frac{v_d}{v_c} - 1 \right) \mathrm{exp} \left[ - (\frac{y-h_\eta}{h_s})^2 \right] , 0.1 \right\} & v_d \geq v_c
        \end{cases}
        .
        \label{eq:eta_2}
\end{equation}

This is motivated by more kinetic reconnection studies, scaling effective resistivity with electron drift velocity (above the threshold velocity, up to a maximum value of 0.1). However, this resistivity is also restricted in heights to those around $h_\eta$ and decays with a characteristic length of $h_s$. From $t=t_\eta$ on, the resistivity directly feeds free energy from the B-field losses into accelerated electron energy, via the non-conservative term in the induction equation relating to $\eta \mathbf{J}$, where $\mathbf{J}$ is current. In the self-consistent beam-MHD evolutions, we modify the energy budget such that all the energy which would otherwise be directed into Ohmic heating, gets channeled into particle acceleration at the points meeting the selection criteria \citep{2020RuanFlare}, and then deposits this energy elsewhere using electron beams that follow field lines and encounter denser chromospheric regions. We will redefine these two resistivity regimes to generate the evaporation by particle beams.

The results sections will run through the sequential and cumulative implementation of the changes listed above in the simulations based on the setup described in \citet{2020RuanFlare, 2023DruettAMRVAC}.

\section{Results}

\subsection{Spatial restriction of anomalous resistivity} \label{sec:eta_spatial}

The anomalous resistivity in the second phase of the scheme presented in \citet{2020RuanFlare} has a decaying dependence with vertical distance from the x-point reconnection site, however, this is supposed to model the acceleration via direct current due to small scale instabilities. This in not the only possible acceleration mechanism, shock acceleration and turbulent reconnection are viable alternatives. Moreover, electron acceleration is not necessarily restricted to the reconnection x-point region, and can also occur in the flare looptops \citep[e.g.][]{1994Masuda, 2022Fleishman}.
Additionally, the reconnection region varies in height over time as the simulation evolves. Therefore, we re-run the experiment with this exponential decay in resistivity with height away from the initial reconnection point removed, and base the resistivity (hence the particle acceleration energy) purely on the value of the ratio of the electron drift velocity compared to the threshold velocity,

\begin{equation}
    \eta (x,y,t \geq t_\eta) =
        \begin{cases}
            0 & v_d > v_c \\
            \mathrm{min} \left\{ \alpha \left( \frac{v_d}{v_c} - 1 \right), 0.1 \right\} & v_d \geq v_c
        \end{cases}
        .
        \label{eq:eta_3}
\end{equation}

\subsection{Beam electron model activation time} \label{sec:eta_timing}

\begin{figure}[!ht]
    \centering
    \includegraphics[width=\textwidth]{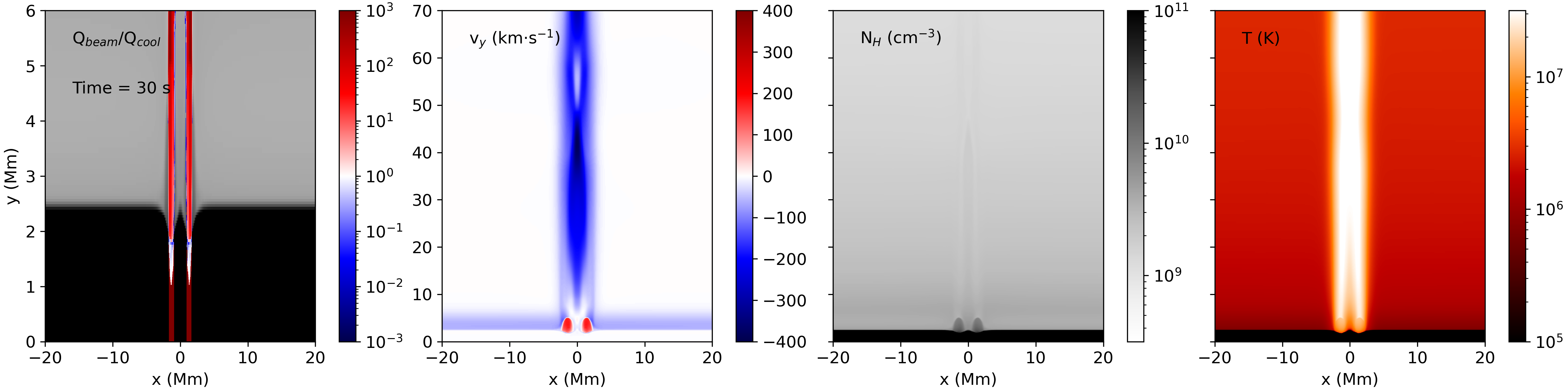}
    \\
    \includegraphics[width=\textwidth]{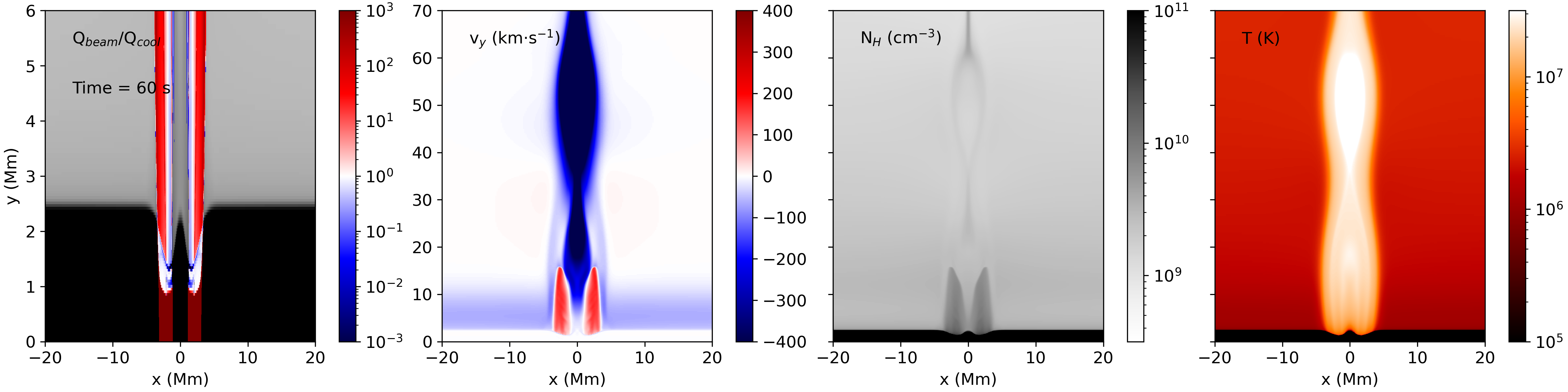}
    \\
    \includegraphics[width=\textwidth]{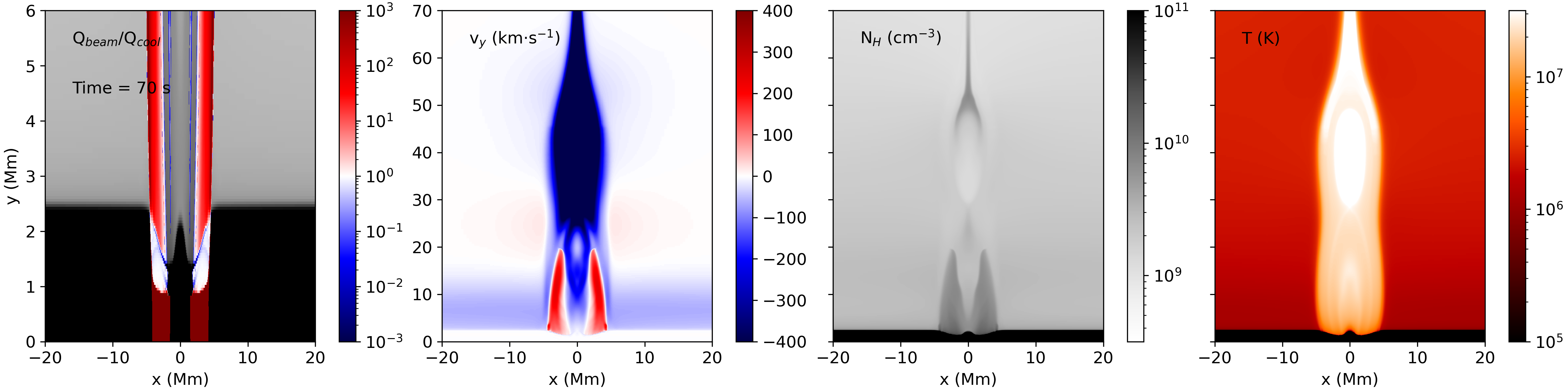}
    \\
    \includegraphics[width=\textwidth]{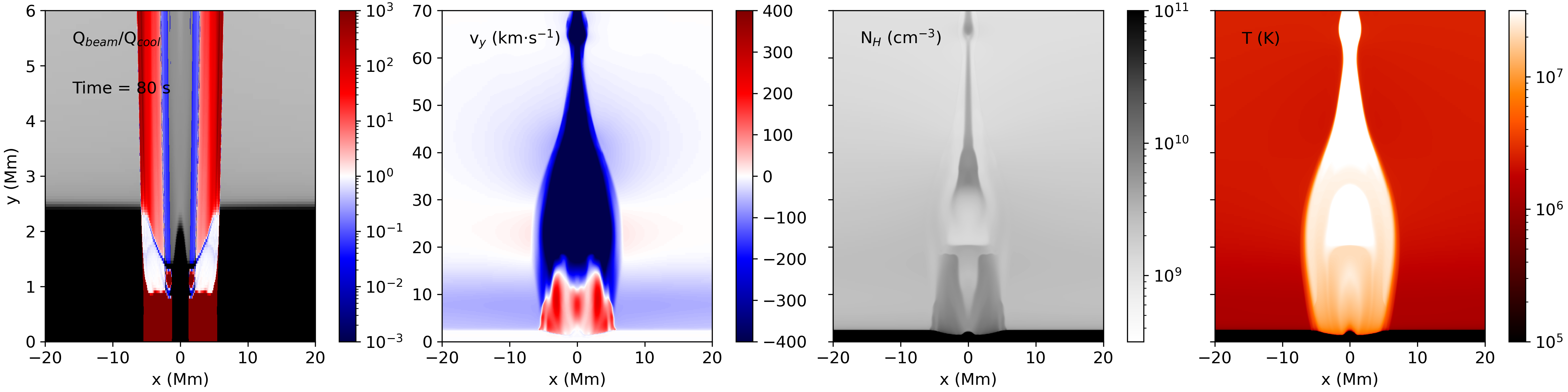}
    \caption{Chromospheric evaporation in a flare  simulation with earlier beam electron activation and spatially de-restricted anomalous resistivity. The left column shows the ratio of beam heating to radiative cooling in red-to-blue, over a background image of the logarithmic number density. The other columns of panels (moving to the right) show zoomed-out views in the y direction with plots of the vertical velocity, plasma number density, and plasma temperature. The top row shows the atmosphere after 30 seconds. Moving downward the subsequent rows show the panels at $t =60$~s, $t =70$~s, and $t =80$~s. An online animated version of this figure is available.}
    \label{fig:eta_spatial_timing}
\end{figure}

The switching between resistivity regimes in previous works \citep{2020RuanFlare, 2023DruettAMRVAC} occurred at $t_\eta=0.4$ experimental time units ($t_\eta=31.2$~s in solar time), however the initially diffuse current sheet had narrowed a long time before this, so a large amount of Ohmic heating has occurred before the electron acceleration modelling was switched on. This preferentially drives the outflow jets over accelerating the electrons. These outflows then reached the chromosphere only seconds after the beam electrons are first accelerated. The electron acceleration should be initiated much earlier in the process, as soon as the current sheet thins and the electron drift velocities exceed the threshold velocity. This will give time for the fast travelling electrons to effect the lower atmosphere before the reconnection outflow jet arrives.

Figure \ref{fig:eta_spatial_timing} shows the results of the simulation with $B_0 = 50$~G as the background magnetic field strength, as in \cite{2023DruettAMRVAC}, but with $\eta$ defined by Equation \ref{eq:eta_3}, with correspondingly reduced switching time $t_\eta = 2$~s.

\begin{figure}[!ht]
    \centering
    \includegraphics[width=\textwidth]{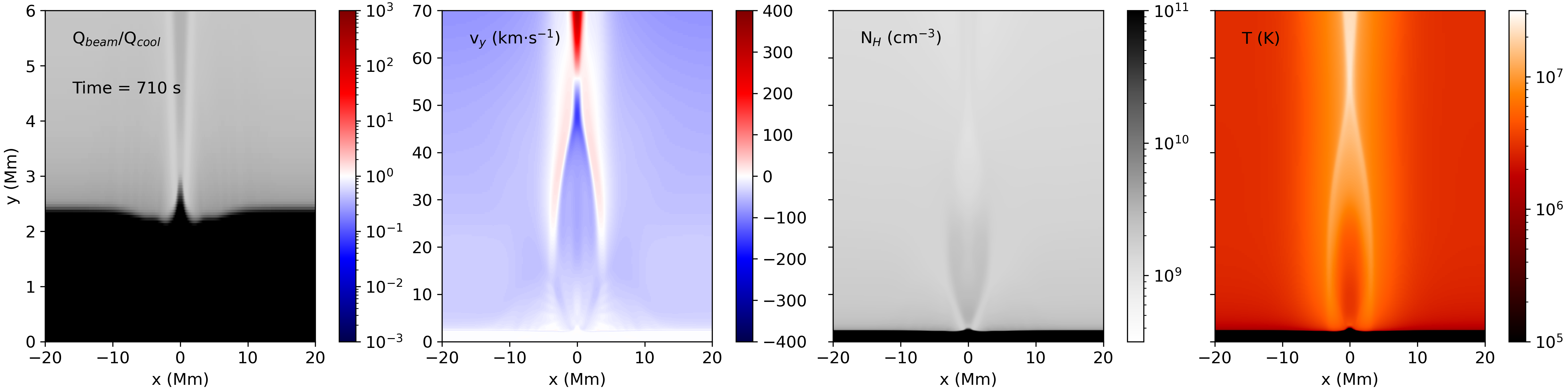}
    \\
    \includegraphics[width=\textwidth]{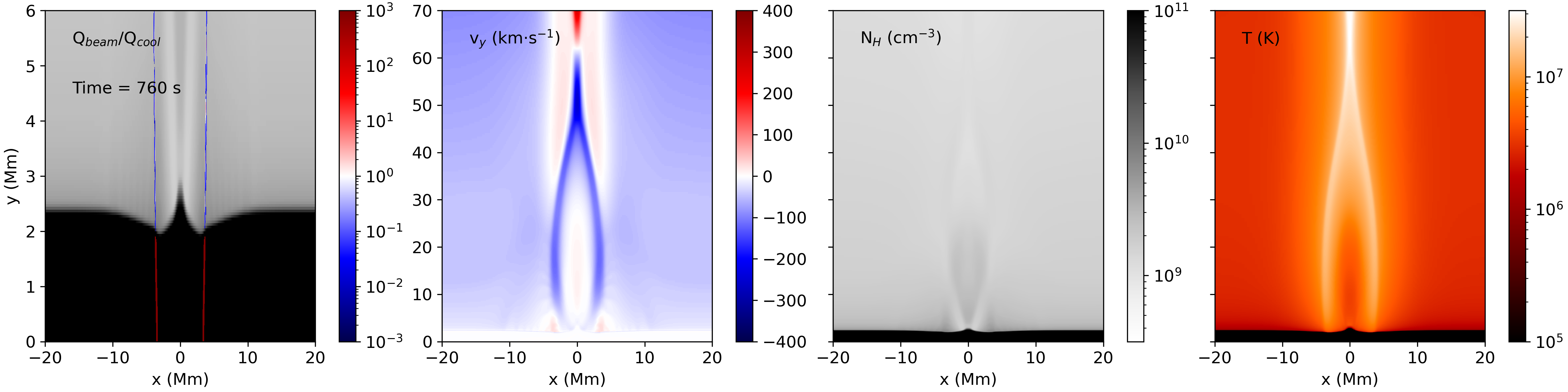}
    \\
    \includegraphics[width=\textwidth]{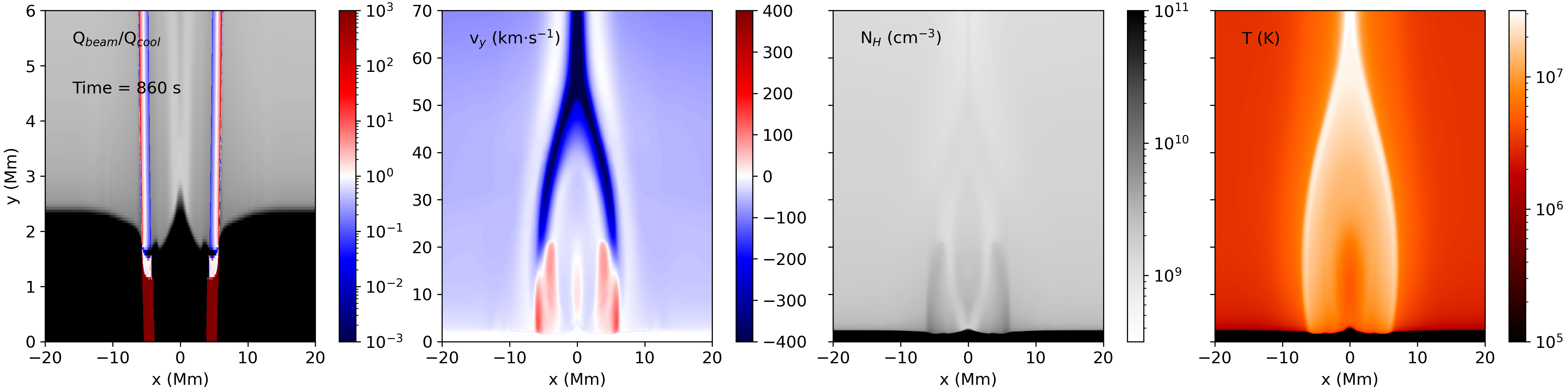}
    \\
    \includegraphics[width=\textwidth]{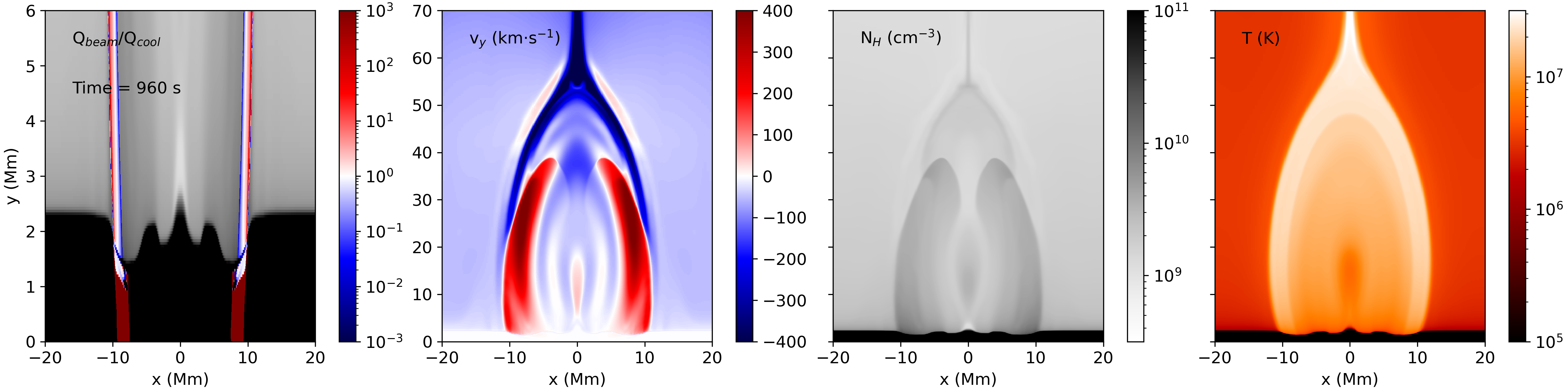}
    \\
    \includegraphics[width=\textwidth]{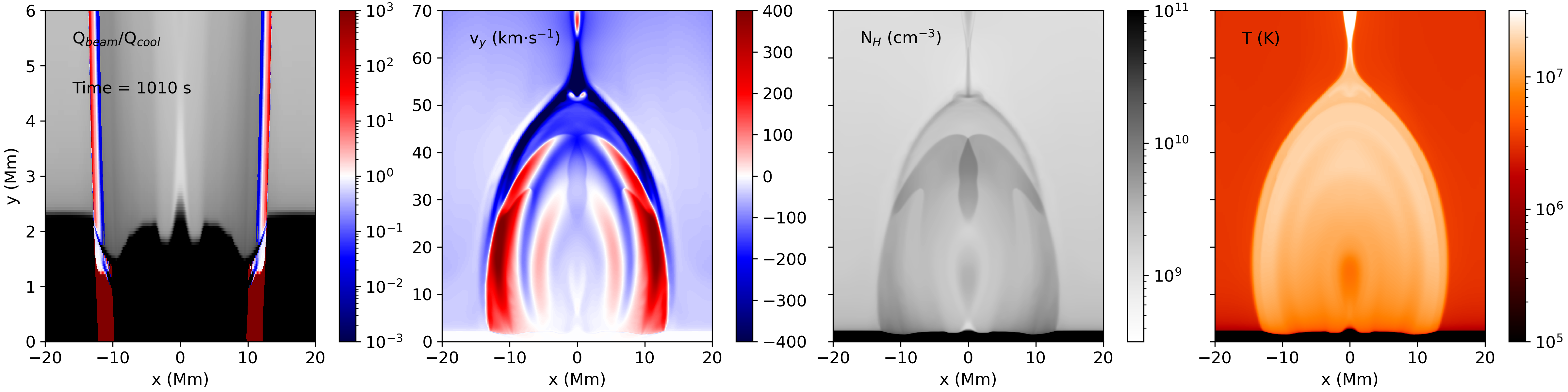}
    \\
    \caption{Chromospheric evaporation in a flare with a gentle precursor phase that pre-forms a narrow arcade. The left column shows the ratio of beam heating to radiative cooling in red-to-blue, over a background image of the logarithmic number density of the base of the model. The other columns of panels (moving to the right) show zoomed-out views in the y direction with plots of the vertical velocity, plasma number density, and plasma temperature. The top row shows the atmosphere 50 seconds before the first electrons trigger the drift velocity criterion stated in Equation \ref{eq:eta_3} ($t =710$~s). Moving downward the subsequent rows show the panels at the time when the first high energy electrons are accelerated ($t =760$~s), 100~s later ($t =860$~s), 200~s later ($t =960$~s), and 250~s later ($t =1010$~s), when the upflowing evaporation streams have collided at the top of the flare loops. An online animated version of this figure is available.}
    \label{fig:gentle}
\end{figure}

In this simulation chromospheric evaporation without the impact and rebound of a magnetic reconnection outflow jet is achieved as can be seen by the upflows shown by the bright red patches in the vertical velocity plots of Figure \ref{fig:eta_spatial_timing}. The left panel demonstrates that the beam electrons in this simulation continue to heat both the chromosphere and the upflowing plasma faster than radiative cooling can cool the flows (see ratios of beam heating to radiative cooling in the left column of Figure \ref{fig:eta_spatial_timing}). Upflows with number densities around $10^{10}$ particles cm$^{-3}$ (third column) are achieved in line with the number densities of upflows seen in other multi-dimensional flare models \citep{2020RuanFlare} using other evaporation mechanisms. However, the upflows are strongly suppressed by the arrival of the downwards reconnection jet which meets the flows just after $t=70$~s. This can be seen in the second column of Figure \ref{fig:eta_spatial_timing} via the dark blue (downward) vertical velocity signature which accumulates in the lower half of the experiment soon after its start. The downward reconnection jet meets the chromospheric evaporation upflows at heights of around 20~Mm between $t=70$~s (third row) and $t=80$~s (fourth row). Although this first modification succeeded in producing evaporation by particle beams in a multi-dimensional model, we are not aware of observations that suggest such an evaporation is suppressed by a strong downwards reconnection outflow jet \citep[rather, observations suggest that evaporation flows are generally sustained for periods of around 10 minutes,][]{2015GrahamEvap}, and so must make further adjustments to the resistivity scheme to produce a solar-like simulation.

\subsection{Gentle precursor phase} \label{sec:eta_gentle}

\begin{figure}[!ht]
    \centering
    \includegraphics[width=\textwidth]{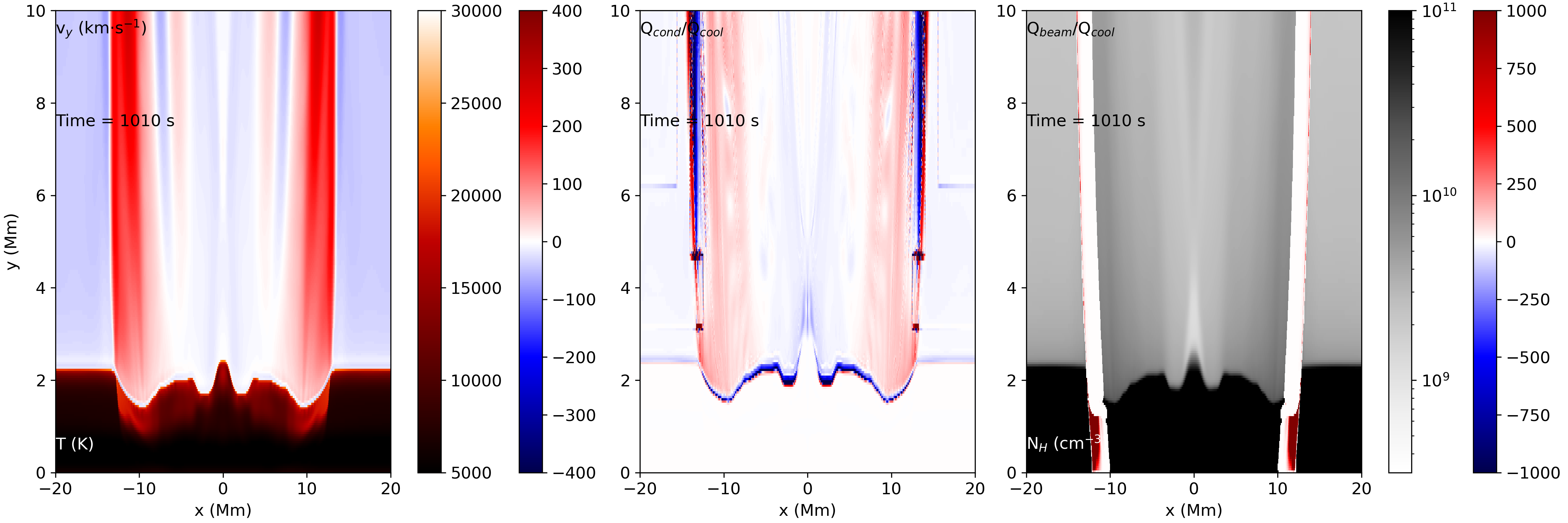}
    \includegraphics[width=\textwidth]{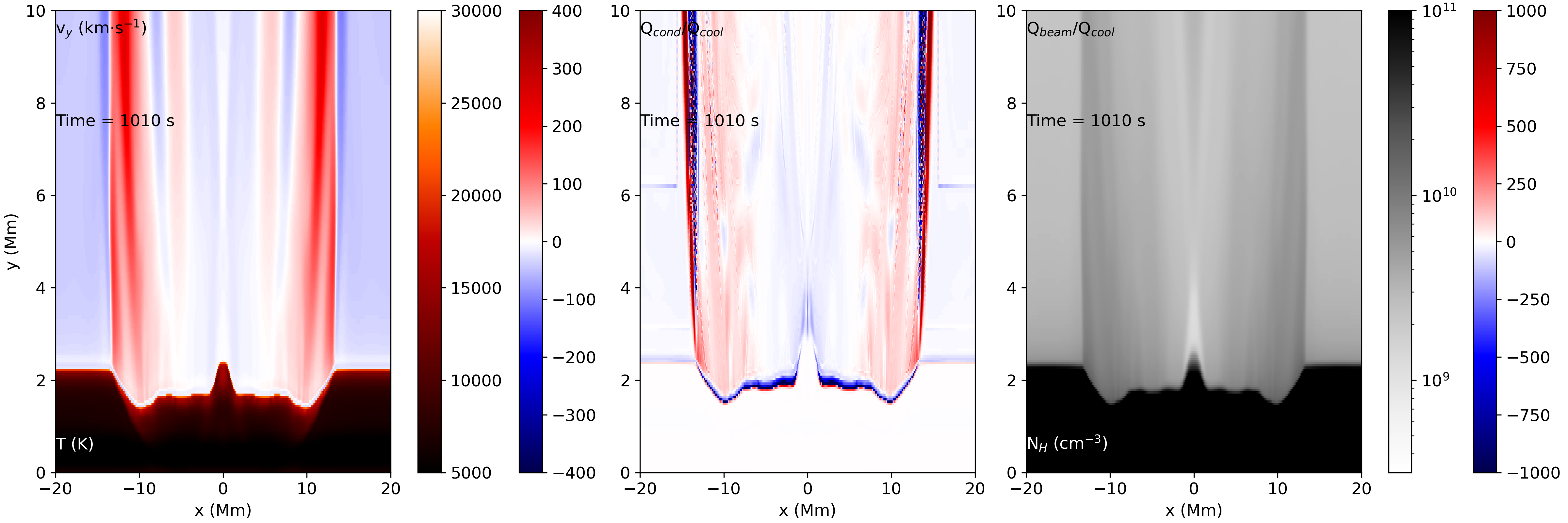}
    \caption{Inspecting the flare evaporation processes and chromosphere. The top panels show the results for the simulation with the beam electrons enabled, the lower panels are the results with the beam energy deposition switched off, but are otherwise identical in magnetic field evolution. Both panels show the conditions at the time $t=1010$~s. The left panel is a composite, the lower region shows the temperature of the atmospheres. Where this saturates to 30000~K the vertical velocities are shown instead. The central column shows the ratio of the thermal conduction to the radiative cooling. The conduction is heating the atmosphere at locations where the colour is red and cooling it where it is blue. The radiative cooling has a small positive number added to it to avoid division by zero in locations where none is modelled. The right column shows the ratio of the beam heating to the radiative cooling, against a background of the plasma number densities in greyscale. The lower right panel shows no beam electrons as they are not activated in this simulation.}
    \label{fig:gentle_evap}
\end{figure}

\begin{figure}[hpt]
    \centering
    \includegraphics[width=0.55\textwidth]{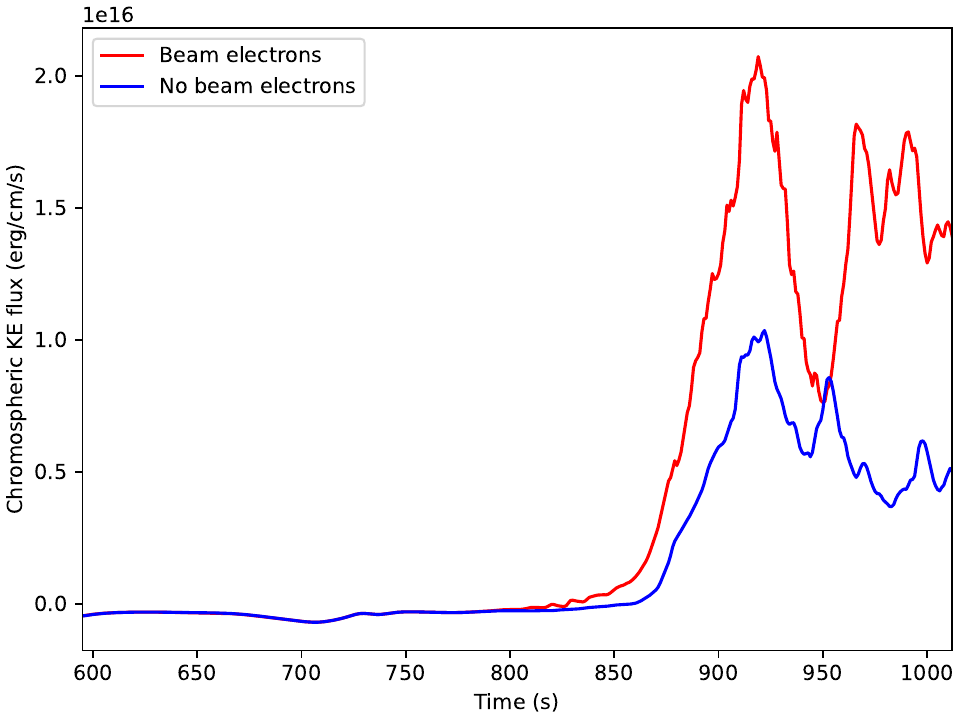} \\
    \includegraphics[width=0.55\textwidth]{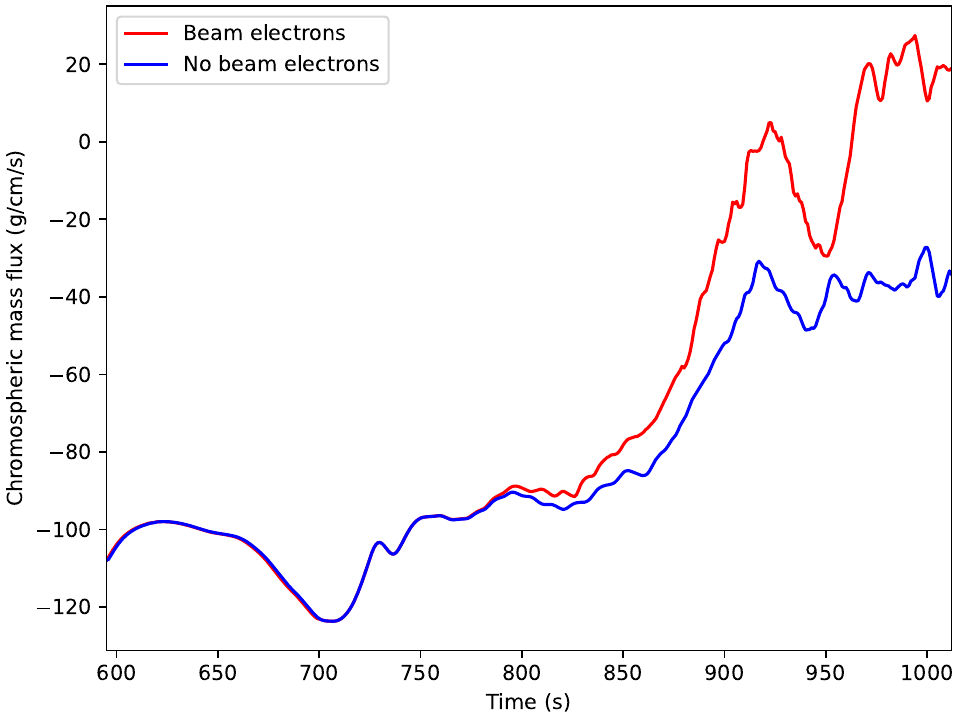} \\
    \includegraphics[width=0.55\textwidth]{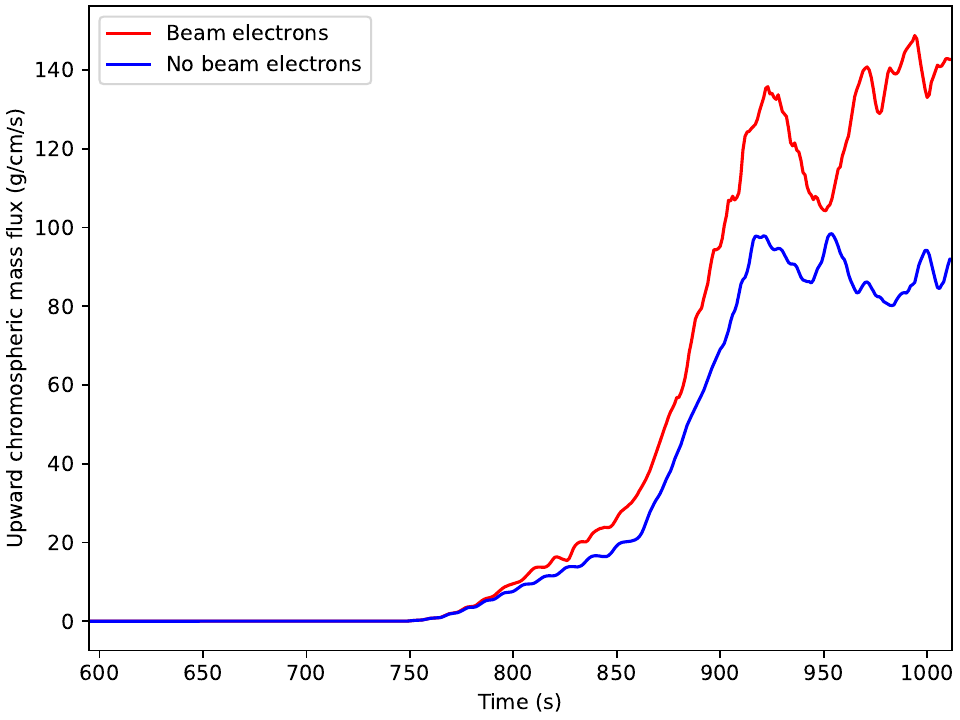}
    \caption{Chromospheric fluxes of kinetic energy and mass as functions of time. These are defined as the fluxes through a horizontal plane across the experiment at a height of 5~Mm. Note that these fluxes are defined in units s$^{-1}$ cm$^{-1}$. The cm$^{-1}$ highlights the 2.5D nature of this simulation. If the flare arch system was assumed to extend 20~Mm ($2 \times 10^9$~cm in cgs) in a direction orthogonal to the 2D panels shown, then to get the fluxes in units s$^{-1}$, one would multiply the numbers on the flux axes by $2 \times 10^9$. These fluxes are shown for experiments with the beam electrons switched on (red line) and otherwise identical evolution but without the influence of the beams (blue line). The top panel shows the kinetic energy fluxes, the central panel shows the chromospheric mass fluxes, and the bottom panel shows the upward mass fluxes, considering only those locations where mass is rising through the 5~Mm height threshold. The electron acceleration mechanism is switched on after $t=600~s$, but the first electrons that meet the drift velocity conditions and are accelerated to deposit their energy in the chromosphere occur at around $t=760~s$.}
    \label{fig:gentle_fluxes}
\end{figure}
\begin{figure}[hpt]
    \centering
    \includegraphics[width=\textwidth]{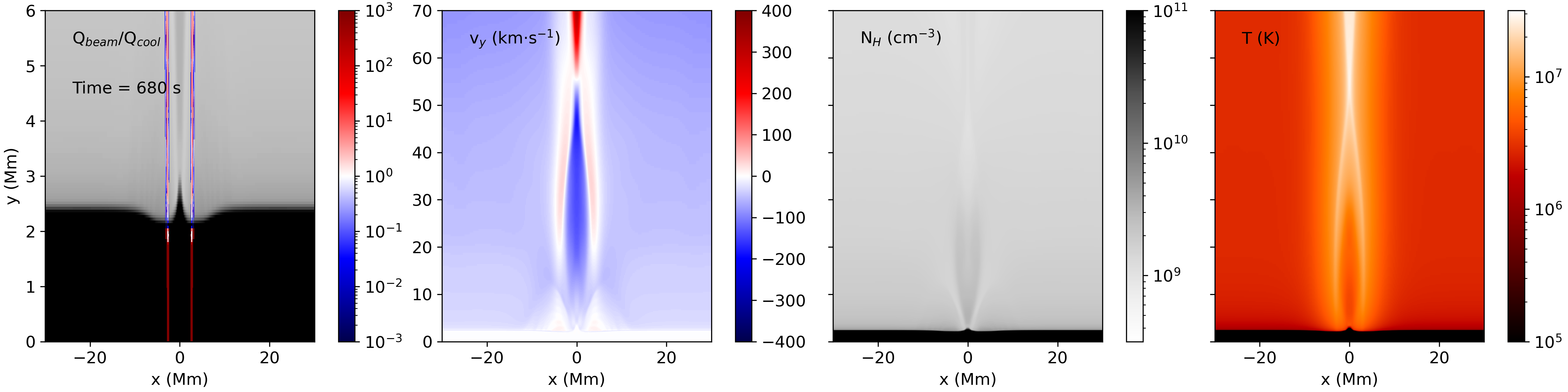}
    \\
    \includegraphics[width=\textwidth]{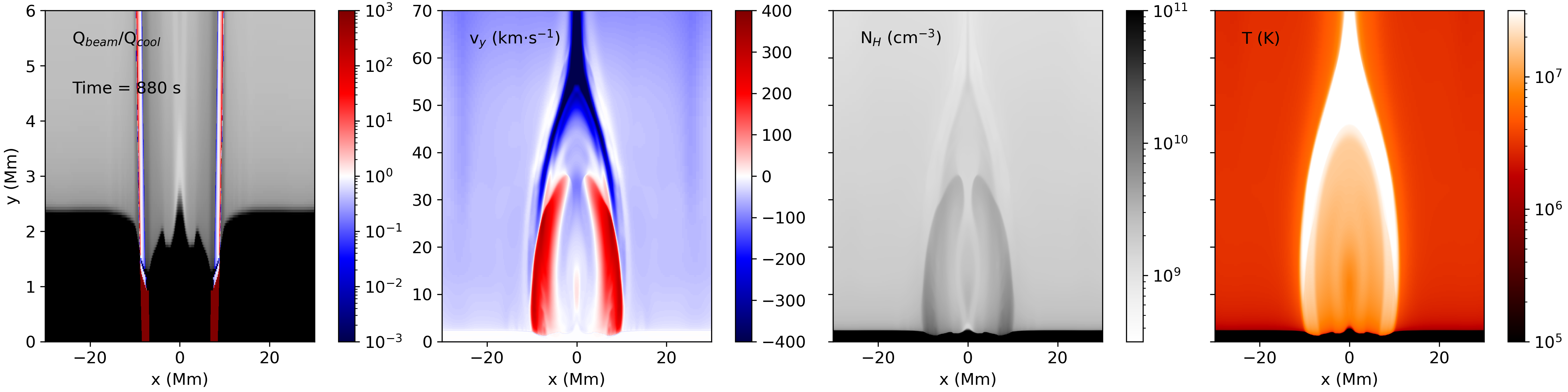}
    \\
    \includegraphics[width=\textwidth]{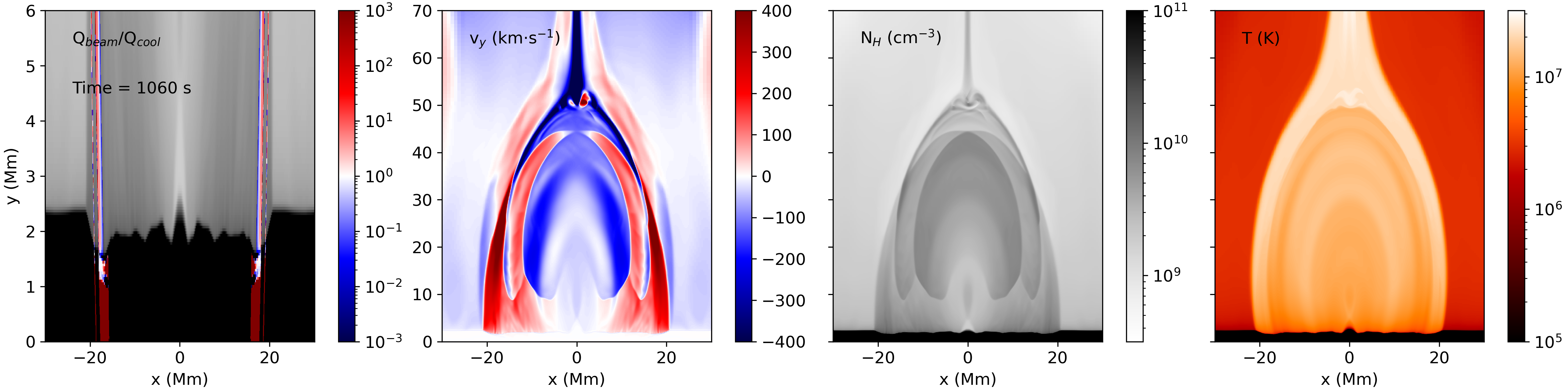}
    \\
    \includegraphics[width=\textwidth]{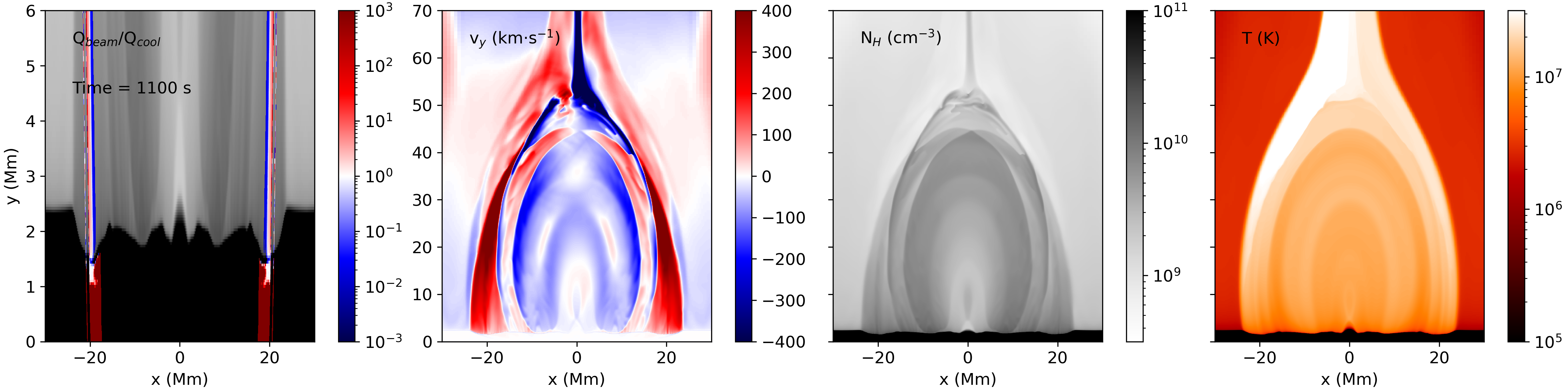}
    \\
    \includegraphics[width=\textwidth]{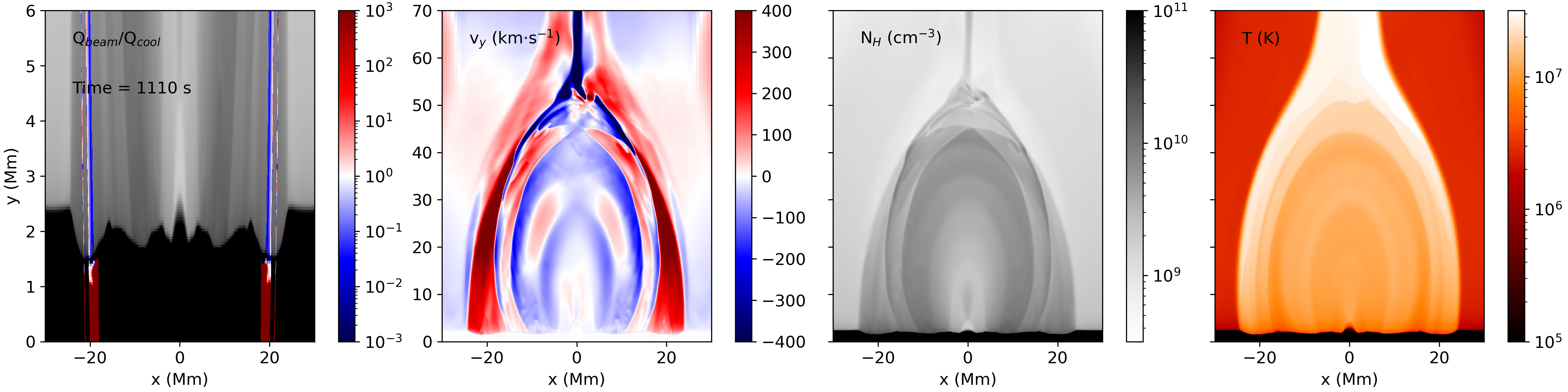}
    \\
    \caption{Chromospheric evaporation in a flare with a gentle precursor phase that pre-forms a narrow arcade, and double the resistivity in Eq.~(\ref{eq:eta_3}). The left column shows the ratio of beam heating to radiative cooling in red-to-blue, over a background image of the logarithmic number density of the base of the model. The other columns of panels (moving to the right) show zoomed-out views with plots of the vertical velocity, plasma number density, and plasma temperature. The top row shows the atmosphere when the first electron beams activate ($t =680$~s). Subsequent rows show the panels 200~s later ($t =880$~s), and 380~s later ($t =1010$~s) when the upward travelling evaporation fronts have met and continued over the looptops to travel down towards the opposite footpoints. The bottom two rows show later times separated by 10 seconds ($t =1100$~s and $t =1110$~s) to demonstrate the later evolution of the flare and the timescales of turbulent motions in the termination shock and looptops. An online animated version of this figure is available.}
    \label{fig:gentle_etaup}
\end{figure}
The direct impact of the reconnection outflow jets on the lower atmosphere is a much more significant agent for chromospheric evaporation than other sources in the experiments of \cite{2001Yokoyama, 2015Takasao, 2020RuanFlare, 2023DruettAMRVAC}, and counters or swamps the evaporation caused by thermal conduction and beam electrons in the examples from Figures \ref{fig:KEsignatures} and \ref{fig:eta_spatial_timing}.

Reconnection outflow jets are legitimate mechanisms to consider for chromospheric evaporation, but are not easily generated in 1D models, highlighting again the urgency for multi-dimensional flare models in studies of chromospheric evaporation. In future works we will investigate observational signatures to help discern between this outflow-impact process as a primary mechanism of evaporation, compared with evaporation via beam electrons and thermal conduction.

If a small loop system is already present below the reconnection point before the outflow jets form, this should focus the reconnection outflow jet's energy into the looptop termination-shock region, preventing the dramatic suppression of the upflows seen through the experiments in Section \ref{sec:eta_timing}.

This scenario can be thought of as equivalent to one in which an arcade is already present due to gentle reconnection or a previous flare (see Figure \ref{fig:gentle}, top row). Gentle precursor phases of flares are often reported and we therefore propose a second modification to our model in this section, which is strongly reminiscent of commonly observed flare evolution scenarios \citep{2007Chifor, 2021HudsonOnset, druett_cocoplot_2022} \citep[also][and references therein]{2023Kontogiannis}. To achieve this we insert a gentle phase with resistivity, $\eta$ of the form given in Equation \ref{eq:eta_1}, but with $\eta_0 = 5 \times 10^{-3}$ instead of the default value $\eta_0 = 3 \times 10^{-2}$ and adjust the thickness of the initial current sheet to be a factor 4 more diffuse than described in \cite{2020RuanFlare}. This gentle phase is run for 598 seconds, and the electrons are switched on after 600 seconds, with resistivity changing at this time to the form given in Equation \ref{eq:eta_3}. However, now the reconnection rate is slowed significantly at this phase, and so there are no regions satisfying the criteria for electron acceleration until 760 seconds. Thus the electrons are now switched on naturally, by the first instance of the threshold velocity being triggered, rather than arbitrarily at a time decided by the experimenter.

Before $t=760$~s we see that that a central downflow has built a small ridge between the two opposite polarity patches either side of it, and thermal conduction has acted to evaporate some small amount of material on either side (Figure \ref{fig:gentle}, top two rows). Once the beam electrons are triggered by their drift velocity exceeding the threshold values, they work in tandem with thermal conduction from the hot flare loops to evaporate plasma continuously over the remaining time of the experiment. As reported in \cite{2023DruettAMRVAC}, the beam electrons heat the material throughout the chromosphere, and the temperature plots confirm that these beam electrons increase the temperatures at much greater depths in the chromosphere than was the case for the simulation with only thermal conduction, and beam electrons switched off (see Figure \ref{fig:gentle_evap}). At these times, the outer footpoints of the beam electrons are well aligned with the edges of the advancing flare ribbons, which is not always the case for multi-dimensional models including energetic beam electrons \citep{2023DruettAMRVAC}. Over the first 250 seconds after the beams are switched on the outer footpoints sweep out from $x =\pm 2.5$~Mm to $x =\pm 12.5$~Mm, advancing horizontally at an average speed of 40 km s$^{-1}$. The upflowing evaporation reaches temperatures up to 10~MK, with higher temperatures (up to 30~MK) still seen in the reconnection outflows and termination shock, as reported in \cite{2015Takasao, 2020RuanFlare}. Evaporating plasma number densities are on the order of $10^{10}$ particles cm$^{-3}$, and higher values when the streams collide at the loop tops. Vertical speeds of up to 500 km s$^{-1}$ occur in the evaporation upflows at heights of around 10-30 Mm. It will be of interest to study future setups with deliberately broken left-right symmetries about the PIL, to assess the possibility of loop-top KH turbulence by interacting evaporation flows, as studied in isolation in \cite{2018RuanKHI}.

To confirm the significance of the electron beams in this simulation, the experiment was re-run with identical conditions, except for switching off the electron energy deposition. Figure \ref{fig:gentle_evap} shows the bases of the models under these conditions, and they appear broadly similar. This simply reiterates that thermal flares are a plausible source of chromospheric evaporation, as in previous studies mentioned in Section \ref{sec:1d}. There are differences, however, which reveal the influence of the electron beams, firstly the upflow speeds are higher, secondly the chromospheric-to-transition-region boundary has significantly different morphology, and thirdly the chromosphere shows much greater heating at deeper depths in the version including beam electrons. Observations of spectral emission from plasma at these depths could thus distinguish the signatures of locations with evaporation likely to be driven by beam electrons compared to those which may be driven by thermal conduction at the top of the chromosphere, or impact and reflection of a reconnection outflow jet. A study comparing lines due to beam driven and thermal conduction driven microflares in 1D has already yielded promising diagnostics, such as the behaviour of the Mg II triplet \citep{2020Testa}.

To quantify the impact of the beam electrons, plots of the kinetic energy flux against time through a horizontal plane at a height of 5~Mm are shown in the upper panel of Figure~\ref{fig:gentle_fluxes}. This confirms that the beam electrons in these simulations more than doubles the net upward flux of kinetic energy from the chromosphere. The upward mass flux is also significantly increased, as shown in the central panel. The net upward mass flux is positive in the case including beam electrons, which was not the case for the weaker flares even via the reconnection outflow impact and reflection process, as demonstrated in \cite{2023DruettAMRVAC}. However, to remove the influence of the downward mass flux, the lower panel shows only the mass flux total from those regions with upward velocities. Again we see a significant enhancement due to the beam electrons with a 56\% increase on the upward mass flux when the beam electrons are enabled. All of these flows are stably maintained by the influence of the beam electrons accelerated due to the continuing reconnection at the x-point, and thermal conduction at the top of the chromosphere.

\subsection{Anomalous resistivity maximum value} \label{sec:eta_maxval}

The maximal anomalous resistivity magnitude is a third free parameter within the simulations, where we used the form given in Equation \ref{eq:eta_3}, with $\alpha=1 \times 10^{-4}$ as in \cite{2020RuanFlare, 2023DruettAMRVAC}. We now re-run the experiment with $\alpha=2 \times 10^{-4}$ and the maximum value changed from 0.1 to 0.2 in order to inspect the effect this has on the flare atmosphere. The experiment evolution is shown in Figure \ref{fig:gentle_etaup}.

The reconnection process is triggered more swiftly in this case, with the first electrons achieving a drift velocity above the particle-streaming threshold at $t=680$~s instead of $t=760$~s (Figure \ref{fig:gentle_etaup}, top row). However, after this the evaporation scenario proceeds relatively similarly to the case with lower resistivity values. As before, 200~s later ($t=880$~s, second row of panels) the evaporation flows are on the verge of meeting at the top of the loops between 30 and 40~Mm above the photosphere. This indicates that the evaporation results are robust to alterations in the resistivity value. A study of the variations of free parameters such as background magnetic field strength in these experiments will be presented in a forthcoming paper, and such a study for evaporation driven by other mechanisms than beam electrons has already been submitted \citep{2023DruettAMRVAC}.

We now trace the full evolution of the loop system over this impulsive phase. The upflows collide and a discernible waveform continues, passing over the looptops, and travelling down to the other side of the loop system (Figure \ref{fig:gentle_etaup}, third row). At around the same time a termination shock forms at the top of the loop system from the reconnection jet outflow. This is achieved in a more gentle fashion than in the experiments of \cite{2001Yokoyama, 2015Takasao, 2020RuanFlare, 2023DruettAMRVAC}, demonstrating clear differences in these multi-dimensional flares with a gentle precursor phase or pre-formed system of arches at their base. 

Finally, dense ($N_H >10^{10}$ cm$^{-3}$) hot ($\sim 10$~MK) flare loops with evaporation speeds reaching $\sim 600$~km s$^{-1}$ at heights of 20-40~Mm are fully filling the postflare loop area from footpoint to footpoint. This has occurred 7 minutes after the energetic electrons were initiated (Figure \ref{fig:gentle_etaup}, bottom two rows).

Two other details can be noted. As was the case for strong flares in \cite{2023DruettAMRVAC}, perpendicular energy transport near the x-point reconnection region is transported down yet-to-reconnect field loops. So although the energetic electron footpoints still demarcate the footpoints of the separatrix between reconnected and yet-to-reconnect magnetic field loops, the hot flare ribbon footpoints extend further outward at later times in the flare, forming pre-heated loops outside of the footpoints of the beam electrons. 
Secondly, looptop turbulence forms near the termination shock of the reconnection outflow within a region of noticeably higher temperature. We see that turbulent eddies seem to oscillate on both sides the polarity inversion line, heating each side of the supra-arcade flows alternately, to temperatures in excess of 30~MK (see Figure \ref{fig:gentle_etaup}, bottom two rows at heights around 60~Mm either side of the line $x=0$). 

This mechanism is known as the ``magnetic tuning fork" model, suggested by \citet{2016TakasaoQPPTuningFork} based on the simulations first presented in \citet{2015Takasao} with a recent analysis of this process in a full 3D flare simulations performed by \citet{2023shibataALT}. It is one of the plausible mechanisms for generating quasi periodic pulsations (QPPs) of solar and stellar flares \citep[See][and references therein]{2018McLaughlinQPP, 2021ZimovetsQPP}. \citet{2018McLaughlinQPP} classify this process under the heading of ``Oscillatory processes of the emitting plasma" within their three categories of QPP generation mechanisms.

\section{Discussion}

We now compare our results briefly to those of previous 1D and multidimensional models, as well as observations.

The number densities of the chromospheric evaporation in the models presented here ($N_H \approx 10^{10}$ cm$^{-3}$) are significantly lower than those of flares produced using RADYN \citep{2005Allred, 2017SimoesFlareContinuum} in which the evaporation flows settle to densities of around $\rho \approx 10^{-11}$ g cm$^{-3}$ or number densities $N_H \approx 10^{12}$ cm$^{-3}$, with electron number densities of $N_e \approx 10^{11}$ cm$^{-3}$ for F10 and F11 flare simulations. ``F10" refers to the exponent of the energy flux of beam electrons at the top boundary, i.e. F10 has beam flux $10^{10}$ erg cm$^{-2}$ s$^{-1}$ and F11 has $10^{11}$ erg cm$^{-2}$ s$^{-1}$. It should be noted that the RADYN F10 results of \citet{2005Allred} were achieved with a constant beam heating along the same field line over several hundred seconds, which was not the case in our models because the reconnecting fieldlines, and hence the fieldlines that have electron beams vary over time. However, \citet{2018Polito} found that the upflow densities in RADYN models were strongly affected by the initial conditions such as loop temperatures, so it can still be possible to generate higher evaporation densities in RADYN models without a longer duration of beam electron heating. The number densities of chromospheric evaporation caused by a 10 second impulse of beam electrons in HYDRO2GEN are closer to those found in our model, for their F10 and 3F10 model with $N_H \approx 10^{10}$ cm$^{-3}$, dropping to $N_H \approx 10^{9}$ cm$^{-3}$ at the tops of the evaporation fronts, and the F11 flare between $N_H \approx 10^{10}$ cm$^{-3}$ and $N_H \approx 10^{11}$ cm$^{-3}$. The upflow velocities of these evaporations are in the range $v=200$~km~s$^{-1}$ to $v=400$~km~s$^{-1}$ for the F10 model, increasing to $v=500$~km~s$^{-1}$ to $v=1000$~km~s$^{-1}$ for the F11 model, which again is in fair agreement with the results in our simulations.

Both the HYDRO2GEN \citep{2018Druett} and RADYN \citep{2005Allred} present results with significantly lower temperature evaporations than the models presented in this study, with the flare loops reaching up to 10-20~MK, compared to 3-4~MK in the RADYN F10 and F11 simulations in \citet{2005Allred} or 2-10~MK of the HYDRO2GEN F10 and F11 simulations \citep{2018Druett}. It is perhaps not so surprising that the characteristic temperatures are lower in the RADYN models referenced, when one considers the greater plasma densities that they manage to evaporate. Higher temperatures on the order of 10-20~MK are produced in RADYN simulations with (a) lower evaporation densities and longer beam durations of 60~s \citep{2019Polito} or (b) just long beam durations, approaching 1000~s \citep{2018Reep}.

Additionally, all of the 1D models are unable to include effects that are specifically multi-dimensional. Examples of these effects include the released magnetic energy that heats the plasma due to compression and Lorentz forces, thermal conduction of energy across the field lines near the reconnection region, the termination shock region, and turbulence at the top of the loop system. Future studies will be used to calibrate the beam fluxes of specific magnetic field lines with those in 1D models so that better comparisons can be made. Some 1D models may also overestimate evaporation densities because in 1D beam heating increases the gas pressure, which then applies a force term $p \mathbf{v}$ in the field-aligned direction. In 2D and 3D this pressure is also applied in the perpendicular directions, providing less energy to evaporation flows.
The pre-heating of yet-to-reconnect field loops via perpendicular energy transfer in the reconnection region is another process that cannot be easily analysed in 1D flare models. This heat is then conducted efficiently in directions parallel to the fieldlines. Later on in the experiment heat conduction from the reconnection region reaches the footpoints of fieldlines before they have reconnected, and thus the leading edge of the hot flare ribbon can actually precede the reconnecting fieldlines at the outer edges of the model (see Figure \ref{fig:gentle_etaup}). This multi-dimensional context provides an alternative explanation to that given in \cite{polito_ribbons_2023} of the leading edges of flare ribbons in spectral lines that were interpreted as showing lower evaporation signatures \citep[see][and citations and discussion therein]{polito_ribbons_2023}. Our findings do support the general interpretation of the authors that flare loops with footpoints slightly inside the leading edges of the ribbon harbour the strongest beams of particles, and are associated with the highest evaporation speeds and densities, as can be seen in the animated version of Figure \ref{fig:gentle_etaup} towards the end of the experiment. This phenomenon manifests over scales on the order of a few IRIS pixels, i.e. on the order of 0.5'', which is easily spatially resolved by our simulation, with grid point separations of 48.8~km at chromospheric heights corresponding to a separation of 0.059'' \citep{2020RuanFlare}. This phenomenon of the beam electron sites being located somewhat inside the leading edges of the flare ribbon will also be addressed for stronger flares in \cite{2023DruettAMRVAC}.

Our simulated chromospheric evaporation by beam electrons is in the regime of ``explosive evaporation" when compared to reports of observations of upflows with speeds of $238$~km~s$^{-1}$ in \Fexix~(592.23~\AA) by \cite{2006MilliganExplosive}, for a flare with a beam electron flux estimated as $\sim$4F10 based on observations with RHESSI \citep{2002LinRHESSI}, and flows of $>200$~km~s$^{-1}$ in \Fexxiii~and \Fexxiv~for a flare estimated to have beam electron fluxes $\sim$5F10 \citep{2009MilliganEvap} as well as other observations with similar velocities \citep{2015GrahamEvap, 2015Tian, 2016Polito}. Gentle evaporation was described to be of the order 110 km~s$^{-1}$ in the \Fexix~line, which is sensitive to temperatures peaking at 8~MK.

Chromospheric evaporation upflows are generally reported in lines that are sensitive to temperatures in the 0.5-10~MK range, but SXR observations of looptops place the characteristic temperatures of those regions to average around 15~MK, observations of blue-shifts have been reported in high-resolution observations \citep[see Section 5.1.1 of ][and references therein]{2021DePontieuIRIS} and in hotter lines such as \Fexxiv~\citep[$\sim20$~MK, ][]{2022Sellers}, although they are generally dominated by stationary components \citep{2009MilliganEvap}. This is in agreement with the patterns of temperatures seen in our models for the plasma from the chromospheric evaporation upflows, compared with that around the termination shock and supra-arcade turbulence, which provides an interpretation for the discrepancies between the maximum temperatures in the 1D and multi-dimensional models. It is only the loops that are not passing through regions of looptop turbulence and the termination shock that have maximum temperatures significantly less than 10~MK in our simulations (see Figure \ref{fig:gentle_etaup}). \cite{2016Polito} reported upflowing material with high temperatures $T>10$ MK, at electron number densities in the coronal lines of order $N_e = 10^{10}$~cm$^{-3}$, in line with values obtained for fully ionised plasma in the simulations presented here. The values obtained for the evaporation velocities, temperatures, and densities in our models are thus within the ranges of values reported in observations.

\section{Conclusions}

In this work we report the first chromospheric evaporation by beam electrons in a multi-dimensional simulation of a solar flare. This has been achieved by building on the pure MHD flare models of \cite{2001Yokoyama} that were turned in self-consistent beam-MHD simulations by \cite{2020RuanFlare}. The main novelty is a-physics-motivated adjustment of the anomalous resistivity that influences the beams, to produce a realistic flare evolution scenario in which evaporation is driven upwards from the chromosphere, reaching speeds of over 250~km/s at only 5~Mm above the photosphere. These evaporation flows accelerate upward and reach heights of up to 50~Mm, with temperatures in the 5-20MK range. The flows from opposite footpoints meet at the apexes of the reconnected flare loops, and the shocks pass over the apexes then travel down towards the opposite footpoints of the flare, in contrast to earlier simulations that use different evaporation mechanisms. Over half of the upward kinetic energy flux from the chromosphere is directly attributable to the electron beams, and 35 \% of the mass flux (a 56\% increase).

Thermal conduction remains a relevant complimentary evaporation mechanism in these simulations. We also clarified the role of the reconnection outflow jet impact on the lower atmosphere in previous multi-dimensional models. This should be considered as a potential source of evaporation, but is generally unavailable in 1D modelling due to the dimensional constraints.

This beam-MHD model is the first of its kind, and has great scope follow-up papers, providing 2D and 3D context to studies of evaporation, as well as testing long-held intuitions regarding solar flares, some of which are based on over 50 years of modelling chromospheric evaporation in one dimension. 

There is a need to further improve the model itself, particularly the beam transport model (e.g. basing the mean pitch angles, and energy spectra of the beam upon information from the atmosphere and evolution of the simulation). Indeed, the beam initial pitch angles were not varied here, while \cite{2020RuanFlare} already demonstrated how that can lead to trapping in loop tops leading to HXR emission there. All this can be built on using the knowledge gained from detailed 1D models. Finally the lower atmosphere of these multi-dimensional models is only a crude representation of the chromosphere. It has a very low magnetic field strength ($<100$~G), non-photospheric densities at the lower boundary, and is lacking convection-zone magneto-convection. Importantly for the chromospheric energy balance, and thus the study of chromospheric evaporation, at present these models still lack detailed non-local thermodynamic equilibrium radiative transfer of 1D models, non-equilibrium partial ionisation, and detailed chromospheric structures. Multidimensional flare models including realistic evaporation processes will bring great detail and spatial context to our understanding of flare processes, complimenting the findings of the next generation of observational instrumentation.

%

%

%
\begin{acks}
We acknowledge the helpful input of the anonymous referee for improving the manuscript.
\end{acks}

\begin{authorcontribution}
M.D. developed the alterations of the flare module code as described in the paper, conducted the simulations and analysis, and wrote the paper.
W.R. wrote the underlying flare module, assisted on the developments presented, advised on analysis, and edited the manuscript.
R.K. is the lead developer of the MPI-AMRVAC code in which this simulation is run, advised on the flare module development, and edited the manuscript.
\end{authorcontribution}
\begin{fundinginformation}
M.D. is supported by FWO project G0B4521N.
M.D., W.R. and R.K. also received funding from the European Research Council (ERC) under the European Union Horizon 2020 research and innovation program (grant agreement No. 833251 PROMINENT ERC-ADG 2018).
W.R. was supported by a postdoctoral mandate (PDMT1/21/027) by KU Leuven. 
R.K. is supported by Internal Funds KU Leuven through the project C14/19/089 TRACESpace and an FWO project G0B4521N. 
The computational resources and services used in this work were provided by the VSC (Flemish Supercomputer Center), funded by the Research Foundation Flanders (FWO) and the Flemish Government, department EWI.
\end{fundinginformation}
%
%
%
%

%
%
\bibliographystyle{spr-mp-sola}
\bibliography{refbib}  
%
%
%
%

\end{article} 
\end{document}